\documentclass{Interspeech2024}

\interspeechcameraready

\title{Audio Fingerprinting with Holographic Reduced Representations}

\name[affiliation={}]{Yusuke}{Fujita}
\name[affiliation={}]{Tatsuya}{Komatsu}

\address{
  LY Corporation, Tokyo, Japan
}
\email{yusuke.fujita@lycorp.co.jp}


\usepackage{amsmath}
\usepackage{bm}
\usepackage[linesnumbered,ruled,vlined,commentsnumbered]{algorithm2e}
\usepackage{booktabs}
\usepackage{multirow}
\usepackage[table]{xcolor}
\usepackage{cite}

\DeclareMathOperator*{\argmax}{arg\,max}

\def\*#1{\bm{#1}}

\begin{document}

\maketitle

\begin{abstract}
This paper proposes an audio fingerprinting model with holographic reduced representation (HRR).
The proposed method reduces the number of stored fingerprints, whereas conventional neural audio fingerprinting requires many fingerprints for each audio track to achieve high accuracy and time resolution.
We utilize HRR to aggregate multiple fingerprints into a composite fingerprint via circular convolution and summation, resulting in fewer fingerprints with the same dimensional space as the original.
Our search method efficiently finds a combined fingerprint in which a query fingerprint exists.
Using HRR's inverse operation, it can recover the relative position within a combined fingerprint, retaining the original time resolution.
Experiments show that our method can reduce the number of fingerprints with modest accuracy degradation while maintaining the time resolution, outperforming simple decimation and summation-based aggregation methods.

\end{abstract}

\section{Introduction}

Audio fingerprinting identifies a song within a database using a segment of an audio signal as a query.
Applications of audio fingerprinting include identifying a user's unknown songs from a microphone input, finding duplicated music in a database, and checking copyrights.
Peak-based matching \cite{Wang2003_ISMIR}, which detects peaks in a spectrogram and encodes their relative positions using hash functions, has traditionally been widely used.
Various approaches to extract more discriminative and robust features than spectrogram peaks have been studied \cite{Cano2005_VLSI, Baluja2008_PR, Hon2015_TASLP, Jiang2019, Wu2022_SPL}.
Most approaches use binary hashing functions for efficient search with hamming distance.
Although the hash-based fingerprint is efficient, noise or distortions in the query audio affect the feature extraction performance, leading to incorrect fingerprints.

Neural-network-based fingerprinting methods, which learn to generate robust embeddings against noise, have advanced the field.
Now Playing \cite{Arcas2017_Neurips} uses a neural network trained with a semi-hard triplet loss function, which minimizes the distance between the reference audio segment and their noisy version while maintaining their distances to other audio segments larger.
Neural audio fingerprinting (NAFP) \cite{Chang21_ICASSP} further exploits an advanced contrastive learning framework and extracts fingerprints with small window shifts (e.g., 0.5 sec), leading to better search accuracy while precisely determining the matched position within a song.

In exchange for better accuracy with high time resolution, NAFP requires significantly larger storage than traditional hash-based fingerprinting because the fingerprint is a real-valued vector of hundreds of dimensions.
Hashing-based embedding mappings, such as Locality-Sensitive Hashing (LSH) \cite{Gionis1999_VLDB}, and vector quantization methods like Product Quantization (PQ) \cite{Jegou2011_PAMI} reduce storage and improve computational efficiency by aggregating the similarity calculations of multiple, partially similar embeddings with a query in a single computation.
In particular, PQ is widely used in general maximum inner-product search (MIPS) systems.

However, current approaches to improve MIPS do not reduce the number of fingerprints that must be searched, which could be considered another dimension of efficiency.
If we could represent a group of fingerprints, e.g., in the same audio track, as another {\it composite} fingerprint, the number of fingerprints can be reduced.
Moreover, we could efficiently handle a {\it containment} search query like ``find a group of fingerprints in which a query fingerprint exists.''
Though any simple aggregation operation, such as decimation or summation within a group, could reduce the number of fingerprints, it leads to a loss of accuracy and time resolution.
Our motivation is to find an appropriate aggregation operation that maintains both accuracy and time resolution.

In this paper, we propose a method to reduce the number of fingerprints by utilizing holographic reduced representations (HRRs).
HRR \cite{Plate95_TNN} is a representation of a compositional structure in distributed representations.
With HRR, circular convolution (denoted by $\circledast$) binds two items, and summation integrates the bounded items in the same vector space.
An illustrative example of HRR is shown in \cite{Nickel16_AAAI}.
According to the example, one can compose a sentence like $\*s = red \circledast cat + blue \circledast dog$ to represent the co-existence of a red cat and a blue dog, where $red$, $cat$, $blue$, $dog$ are item vectors in the same dimensional space.
Then, one can retrieve the cat's color with the inverse operation, $\*s \circledast cat^\dagger \approx red$ under some assumptions in the associate vectors.
Our proposed method uses this composition scheme to group a sequence of fingerprints.
Each fingerprint is associated with its relative position in a sequence using circular convolution, and then the results are summed together to obtain a composite fingerprint.
It reduces the number of fingerprints stored in the database, while we can retrieve the original fingerprint location through the inverse operation.

We conducted fingerprint search experiments using the FMA dataset \cite{Defferrard2017_ISMIR}.
We followed a similar setup used in the NAFP paper \cite{Chang21_ICASSP}.
The experimental results show that the proposed method can aggregate fingerprints with a slight accuracy degradation compared with non-reduced fingerprints.
It outperforms simple decimation-based and summation-based aggregation methods, which make it hard to recover the original fingerprint location within a sequence.
Though our proposed method can work with any pretrained fingerprinter such as NAFP, we further explored the possibility of using HRR-aware training for a neural fingerprinter.
A similar training strategy with HRR has been proposed for extreme multi-label classification in \cite{Ganesan21_Neurips}.
We are believed to be among the first to apply HRR-based training to contrastive learning.
These additional experiments exhibit that considering the HRR's noise in the training does not offer a significant improvement.
Finally, we discuss the limitations and future work based on the results of HRR-aware training.





\section{Background}

\subsection{Neural Audio fingerprinting}
\label{sec:nafp}
NAFP \cite{Chang21_ICASSP} has introduced the contrastive learning framework to extract a fingerprint for short audio segments.
A neural-network-based function $f$ transforms $T$-length audio feature sequence $\*A \in \mathbb{R}^{F \times T}$ into an fingerprint $\*x \in \mathbb{R}^{D}$:
\begin{align}
    \*x = f(\*A).
\end{align}
A replica audio is prepared for each audio segment $\*A$ in a training set with various augmentations, and the fingerprint $\*r$ for the replica is produced:
\begin{align}
    \*r = f(\mathsf{Aug}(\*A)).
\end{align}
We train the fingerprint function $f$ with contrastive learning.
Given a training batch of $B$ fingerprints $\*X = [\*x^{(1)},\dots, \*x^{(B)}]$ and their replicas $\*R = [\*r^{(1)},\dots, \*r^{(B)}]$, a contrastive loss is calculated as follows:
\begin{align}
    \mathcal{L}(\*R, \*X) = - \sum_{i=0}^B \log \frac
    {\exp(\mathsf{sim}(\*x^{(i)}, \*r^{(i)}) / \tau)}
    {\sum_{j=1}^{B} \exp(\mathsf{sim}(\*x^{(j)}, \*r^{(i)}) / \tau)}, \label{eq:contrast}
\end{align}
where $\tau$ is a temperature hyperparameter, cosine-similarity is used as a similarity measure $\mathsf{sim}$.
Since this contrastive loss only considers mapping from $\*R$ to $\*X$, we also calculate the loss in the reverse direction to encourage one-to-one correspondence between $\*R$ and $\*X$:
\begin{align}
    \mathcal{L}_\mathsf{NAFP} = (\mathcal{L}(\*R, \*X) + \mathcal{L}(\*R, \*X)) / 2.
\end{align}

\subsection{Holographic reduced representations}
\label{sec:hrr}
HRR \cite{Plate95_TNN} uses circular convolution to {\it bind} two vectors.
\begin{align}
    \*a \circledast \*b &= \mathcal{F}^{-1}(\mathcal{F}(\*a) \mathcal{F}(\*b)),
\end{align}
where $\mathcal{F}$ is the discrete Fourier transform.
Then, summation can {\it bundle} multiple vectors into a single composite vector with the same dimension:
\begin{align}
\*s = \*a \circledast \*b + \*c \circledast \*d \in \mathbb{R}^D.
\end{align}
Assuming that the elements of vectors are i.i.d. with zero mean and variance $1/D$,
which is a reasonable assumption given that contrastive learning ensures uniformity in the learned representations \cite{wang2020understanding},
$\*a$ can be recovered using the following inverse operation:
\begin{align}
   \*{\hat{a}} = \*s \circledast \*b^\dagger &= \mathcal{F}^{-1}(\mathcal{F}(\*s) \mathcal{F}(\frac{1}{\*b})),
\end{align}
where $\*b^\dagger = \mathcal{F}^{-1}(\mathcal{F}(\frac{1}{\*b}))$.
The recovered vector has noise due to other bundled vectors ($\*c$ and $\*d$).
The capacity, the number of acceptable vectors to be bundled, increases linearly as the vector dimension $D$ increases.

Note that we do not need to use the inverse operation to check if the vector is in a composite vector. We can ask if two vectors $\*a$ and $\*b)$ are bounded and bundled in $\*s$ by checking that $(\*a \circledast \*b)^\top \*s \approx 1$.
We use this property for efficient fingerprint search.

\section{Proposed method}

In general fingerprinting methods, such as NAFP, many fingerprints are handled independently.
Contrary to such a trend, we attempt to aggregate a sequence of audio fingerprints utilizing the compositional structure of HRR.
As described in the introduction, HRR can represent a composite sentence ``a red cat and blue dog'' as a vector $\*s = red \circledast cat + blue \circledast dog$.
Similarly, the proposed method considers a structure of fingerprints; in this study, a ``sequence'' structure is encoded using HRR.
We empirically demonstrate that the HRR-based fingerprint enables us to perform the containment search to determine whether the query is bounded in the composite representation of a sequence.

\subsection{Composition of fingerprint sequence with HRR}

\begin{figure}[t]
    \centering
    \includegraphics[width=0.95\linewidth]{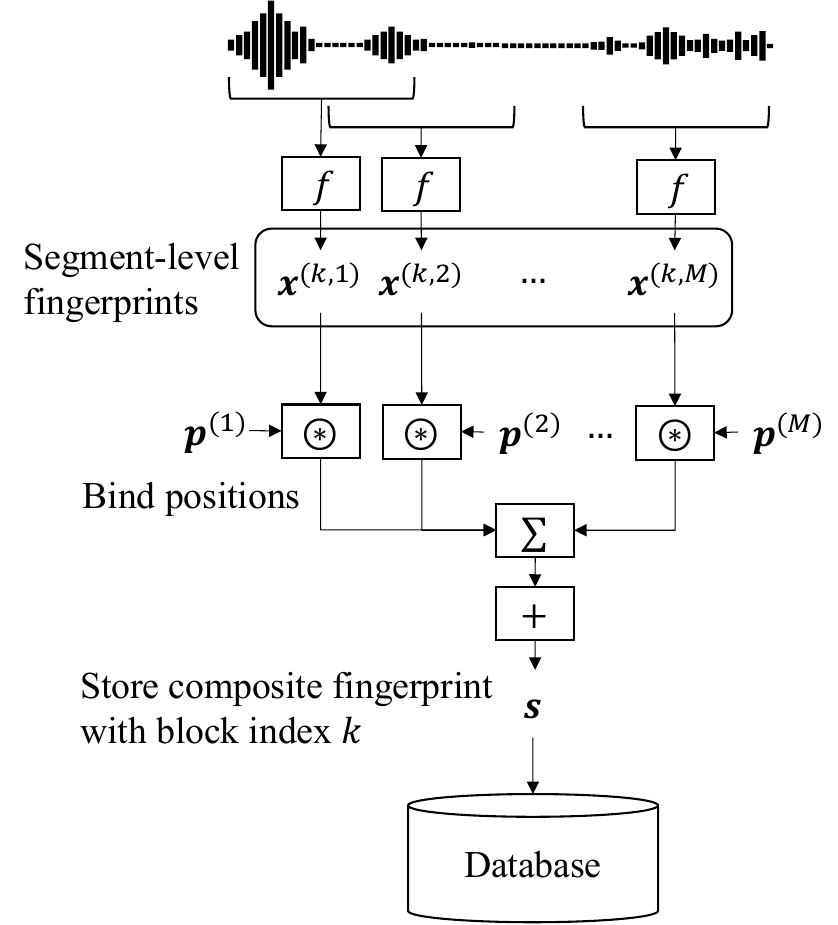}
    \caption{Composition of fingerprints}
    \label{fig:compose}
\end{figure}

The proposed composition method is depicted in Fig. \ref{fig:compose}.

We initialize $M$ position vectors $\*p^{(1)}, \dots, \*p^{(M)} \in \mathbb{R}^D$.
A sequence of $N$ fingerprints $\*x^{(1)}, \dots, \*x^{(N)} \in \mathbb{R}^D$ in the audio database is first segmented into blocks, each with $M$ consecutive fingerprints.
We bind a sequence of fingerprints for each block with the position vectors, resulting in a composite fingerprint $\*s^{(k)}$:
\begin{align}
    \*s^{(k)} = \sum_{m=1}^M \*x^{(k,m)} \circledast \*p^{(m)} \quad (1 \le k \le \lceil N/M \rceil), \label{eq:compose}
\end{align}
where $k$ is a block index and $\*x^{(k,m)} = \*x^{((k-1)M +m)}$ is the $m$-th fingerprint in the $k$-th block.
We only store the composite fingerprint instead of all $M$ fingerprints, which requires $M$ times smaller storage.

\subsection{Search composite fingerprint with positions}

\begin{figure}[t]
    \centering
    \includegraphics[width=0.9\linewidth]{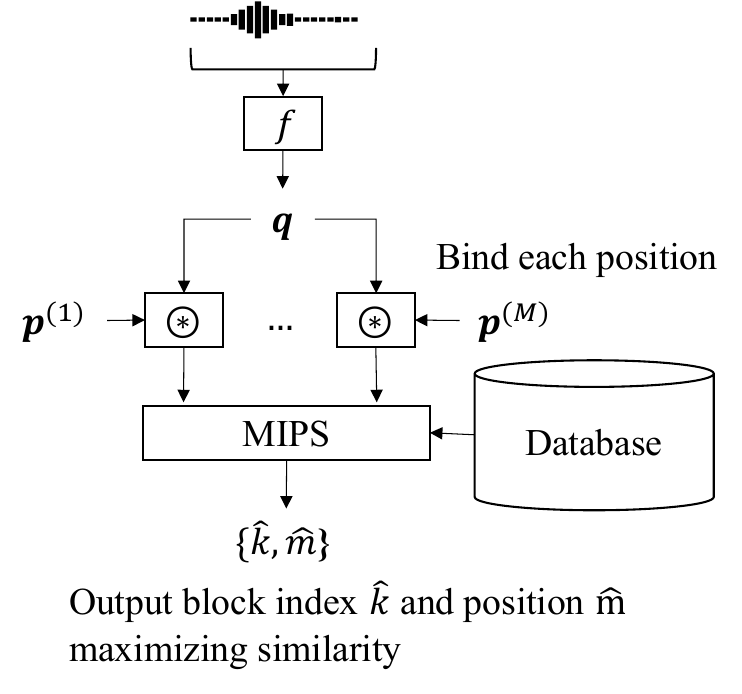}
    \caption{Search composite fingerprint with maximum inner-product search}
    \label{fig:search}
\end{figure}

Fig. \ref{fig:search} shows our search method for composite fingerprints.
Given a fingerprint $\*q \in \mathbb{R}^D$ extracted from a query audio,
we search a block in which $\*q$ exists:
\begin{align}
    \hat{k}, \hat{m} = \argmax_{k,m} \mathsf{sim}(\*q \circledast \*p^{(m)},  \*s^{(k)}) \quad (1 \le m \le M). \label{eq:search_block}
\end{align}
Here, we can use an efficient $K$-nearest neighbor search algorithm for MIPS and produce top-$K$ block indices for each $m$.
Then, we easily obtain a relative position $\hat{m}$ in the retrieved block $\hat{k}$ having the maximum similarity score.

The proposed composition method preserves the distinction between positions, unlike simple decimation or summation operations. We compare the composition operations in the experiment section.

\subsection{Search for a sequence of fingerprints}

When query audio is longer than $T$ (one segment), the system can gather similarity scores for multiple consecutive segments to improve search accuracy.
Given $L$ consecutive query fingerprints $\*q^{(1)}, \dots, \*q^{(L)}$,
we construct a sequence of composite queries as follows:
\begin{equation}
    \*q'^{(i)} = \sum_{m=1}^M \*q^{(i+m-1)} \circledast \*p^{(m)} \quad (1 \le i \le L-M+1).
\end{equation}
Then, we find top-$K$ similar blocks for each $\*q'^{(i)}$.
The offset in the retrieved block index $\hat{k}$ for $i$-th query is compensated by $\hat{k} - i$.
The sequence-level similarity score is the sum of all similarity scores assigned to the same block index, and finally the system outputs the index with the highest score.

\subsection{HRR-aware training of neural fingerprinter}
\label{sec:train_hrr}
Although the proposed method can work with any pretrained fingerprinter, this section further investigates the possibility of learning with HRR's characteristics.

As described in Sec \ref{sec:hrr}, the recovered vectors from HRR have noise, leading to degraded search performance.
To mitigate the issue, we train the fingerprint function $f$ to be aware of the HRR's noise.

Given a training batch of $B$ fingerprints $\*X$ and their replicas $\*R$,
we first generate a batch of composite fingerprints $\*S = [\*s^{(1)},\dots, \*s^{(B/M)}]$ using Eq. \ref{eq:compose}.
The replicas are bounded with position vectors aligned with the batch of composite fingerprints.
A batch of the bounded vectors $\*V = [\*v^{(1)}, \dots, \*v^{(B)}]$ is calculated as:
\begin{align}
    \*v^{(b)} = \*r^{(b)} \circledast \*P_{b\ \mathrm{mod} M} \quad (1 \le b \le B).
\end{align}
Then, we compute the following contrastive loss instead of Eq. \ref{eq:contrast}:
\begin{align}
    \mathcal{L}'(\*V, \*S) &= - \sum_{b=1}^B \log \frac
    {\exp(\mathsf{sim}(\*s^{( \lceil b / M \rceil)}, \*v^{(b)}))}
    {\sum_{k=1}^{B/M} \exp(\mathsf{sim}(\*s^{(k)}, \*v^{(b)}))}
\end{align}
Unlike the original loss $\mathcal{L}$, the modified loss considers noise added to the composite fingerprint and maximizes the similarity only if a query is placed at the specified position.
The loss for mapping from $\*S$ to $\*V$ is slightly different because $M$ consecutive vectors in $\*V$ should be mapped to one composite fingerprint in $\*S$.
To force one-to-one mapping in the loss calculation, we split $V$ according to their positions:
\begin{align}
    \mathcal{L}''(\*S, \*V) = - \sum_{m=1}^M \sum_{k=1}^{B/M} \log \frac
    {\exp(\mathsf{sim}(\*s^{(k)}, \*v^{(k, m)}))}
    {\sum_{k'=1}^{B/M} \exp(\mathsf{sim}(\*s^{(k)}, \*v^{(k', m)})},
\end{align}
where $\*v^{(k,m)} = \*v^{((k-1)M +m)}$.
Then, we mix the two losses:
\begin{align}
    \mathcal{L}_\mathsf{NAPF-HRR} = (\mathcal{L}''(\*S, \*V) + \mathcal{L}''(\*V, \*S)) / 2.
\end{align}

\section{Experimental setup}

\subsection{Data}
We conducted audio fingerprinting experiments on the FMA dataset \cite{Defferrard2017_ISMIR} according to the NAFP paper \cite{Chang21_ICASSP}.
Note that the dataset described below is the {\it mini} version and can be downloaded from \cite{ahym-e477-21}, which is different from the {\it full} version reported in the paper \cite{Chang21_ICASSP}.

The training data set for the fingerprinting function is sampled from \texttt{fma\_medium}, comprising 10,000 songs, each with 30-second audio.
The test-DB dataset is another subset from \texttt{fma\_medium}, comprising 500 songs of 30 seconds each.
The test-query dataset is a noisy copy of the test DB with a random augmentation pipeline, including time offset modulation up to $\pm 200$ ms, background noise mixing using AudioSet \cite{Gemmeke2017_ICASSP} in the SNR range from 0 to 10 dB, and impulse response convolution using two public datasets \cite{Xaudia, Jeub2009_ICDSP}.
The test-dummy-DB dataset is used as a set of distractors; they should not be matched to the test query.
The dummy dataset is sampled from \texttt{fma\_full}, consisting of 10,000 songs, each with 30 seconds.

\subsection{Network architecture and training configurations}

We also used the same network architecture with NAFP \cite{Chang21_ICASSP}.
The network accepts a log-scaled Mel-spectrogram representing 1-second audio with the 0.5-second shift.
The input runs through eight convolutional encoders with separable convolution, layer normalization, and ReLU activation, followed by a projection layer and L2-normalization.
We mainly used the fingerprint dimension $D = 512$, larger than $D=128$ reported in \cite{Chang21_ICASSP}, because our HRR requires a sufficient dimension to bind multiple vectors.
We set the batch size $B$ to 640.
We trained the network using Adam optimizer with an initial learning rate of 1e-4 and cosine decay to 1e-6 in 100K steps.
The temperature $\tau$ was set to 0.05.
We implemented the training pipeline by ourselves with PyTorch.

For training with HRR described in Sec. \ref{sec:train_hrr}, we tested different numbers of positions $M = {2, 4}$

\subsection{Search algorithm}

Faiss \cite{johnson2019billion} is used for efficient MIPS.
We used the inverted file index structure with PQ (IVF-PQ).
For the IVF-PQ, we had 200 centroids with a code size of 64 and 8 bits per index.

\subsection{Evaluation protocol}

We use the Top-1 hit rate (\%) to measure the search performance.
Assuming we have $Q$ query fingerprints and $Q_\mathsf{hit}$ queries with the maximum similarity are hit as top-1,
the Top-1 hit rate can be measured as ${Q_\mathsf{hit}}/{Q}$.

\section{Results and discussion}

\subsection{Comparison of fingerprint aggregation methods}

The proposed aggregation method {\bf HRR} (Eq. \ref{eq:compose}) was compared with two simple alternatives, 1)
{\bf Summation}: $\*s^{(k)} = \sum_{m=1}^M \*x^{(k,m)}$, and 2) {\bf Decimation}: $\*s^{(k)} = \*x^{(k,1)}$.

Table \ref{tab:agg} shows the results on the Top-1 hit rate.
We observed that for all query lengths and block sizes, the proposed HRR outperformed other aggregation methods.
In particular, when query length is 1-sec, i.e., one segment, HRR produced significantly better accuracy than the summation and decimation methods.
Unlike the other methods, it demonstrates that HRR can preserve the original time resolution.
Although we can see significant performance degradation compared with the no-aggregation system, the proposed method can recover the accuracy according to the query length.

\begin{table}[t]
\setlength{\tabcolsep}{5pt}
\caption{Top-1 hit rate (\%) with different aggregation methods. The fingerprint dimension is 512. $M$ is the block size for aggregation.} 
\label{tab:agg}
\centering
\begin{tabular}{ccrrrrrr} \toprule
\multirow{2}{*}{Method} & \multirow{2}{*}{$M$} & \multicolumn{5}{c}{Query length (s)}  \\
 & & 1 & 2 & 3 & 5 & 10 \\ \midrule
\textcolor{gray}{No-aggregation} & \textcolor{gray}{1} & \textcolor{gray}{71.1} & \textcolor{gray}{90.4} & \textcolor{gray}{95.1} & \textcolor{gray}{97.9} & \textcolor{gray}{99.4}  \\ \midrule
Summation & 2 & 31.2 & 74.2 & 84.9 & 92.7 & 96.4 \\
Decimation & 2 & 39.0 & 74.0 & 86.0 & 93.4 & 95.9 \\
HRR (proposed) & 2 & \bf 58.8 &\bf 83.8 &\bf 90.9 &\bf 95.1 &\bf 97.8 \\ \midrule
Summation & 4 &  3.0 & 6.6 & 39.4 & 44.3 & 89.8 \\
Decimation & 4 & 19.3 & 45.8 & 37.7 & 71.3 & 92.8 \\
HRR (proposed) & 4 & \bf 31.0 & \bf 58.3 & \bf 45.0 & \bf 79.0 & \bf 96.0 \\
\bottomrule
\end{tabular}
\end{table}

Table \ref{tab:near} shows the Top-1 {\it near} match rate for the 1-second query.
The near match means that the Top-1 hypothesis is within $\pm 500$ msec.
With $M = 2$, the summation method can produce a better Top-1 near match, suggesting the summation does reasonable aggregation while ignoring the position in a sequence.
With $M = 4$, the summation and decimation methods failed even for near matches.
The proposed HRR method showed no significant difference between the Top-1 exact and Top-1 near values.
It suggests that HRR can accurately discriminate the position in a sequence.

\begin{table}[t]
\setlength{\tabcolsep}{5pt}
\caption{Comparison of Top-1 exact/near match rate for 1-second query with different aggregation methods. The fingerprint dimension is 512. $M$ is the block size for aggregation.} 
\label{tab:near}
\centering
\begin{tabular}{ccrr} \toprule
Method & $M$ & Top-1 exact & Top-1 near \\ \midrule
Summation & 2 & 31.2 & \bf 62.9 \\
Decimation & 2 & 39.0 & 45.0  \\
HRR (proposed) & 2 & \bf 58.8 & 60.4 \\ \midrule
Summation & 4 & 3.0 & 7.7 \\
Decimation & 4 & 19.3 & 22.9  \\
HRR (proposed) & 4 & \bf 31.0 & \bf 32.4 \\
\bottomrule
\end{tabular}
\end{table}

\subsection{Effect of HRR-aware training}

Table \ref{tab:train} shows the results of HRR-aware training.
For $M = 2$, HRR-aware training was slightly better than that without HRR-aware training.
However, the difference was not significant.
For $M = 4$, HRR-aware training was only slightly better at the query length of 10.
We hypothesize that the linear operations of HRR limit the capacity of representations.
Adding some non-linear operations for aggregation could lead to an improvement in the proposed training scheme.
We leave this direction for future work.

\begin{table}[t]
\setlength{\tabcolsep}{4pt}
\caption{Top-1 hit rate (\%) with and without HRR-aware training. The fingerprint dimension is 512. $M$ is the block size for aggregation.} 
\label{tab:train}
\centering
\begin{tabular}{ccrrrrr} \toprule
\multirow{2}{*}{Method} & \multirow{2}{*}{$M$} & \multicolumn{5}{c}{Query length (s)}  \\
 & & 1 & 2 & 3 & 5 & 10 \\ \midrule
HRR & 2 & 58.8 & 83.8 & 90.9 & 95.1 & 97.8 \\
+ HRR-aware train. & 2 & 59.0 & 83.8 & 91.5 & 95.4 & 98.1 \\ \midrule
HRR & 4 & 31.0 & 58.3 & 45.0 & 79.0 & 96.0 \\
+ HRR-aware train. & 4 & 24.4 & 57.0 & 44.5 & 67.4 & 97.0 \\
\bottomrule
\end{tabular}
\end{table}

\section{Conclusion}

We proposed an audio fingerprinting model with holographic reduced representation (HRR).
The proposed method can reduce the number of stored fingerprints by utilizing HRR to aggregate multiple fingerprints into a composite fingerprint.
We conducted fingerprint search experiments using the FMA dataset.
The results show that the proposed method can aggregate fingerprints with a slight accuracy degradation compared with non-reduced fingerprints.
It significantly outperformed decimation-based and summation-based aggregation methods.
While the baseline aggregation methods make it hard to recover the original fingerprint position within a sequence, the proposed HRR-based aggregation successfully preserved it.
Experiments with the HRR-aware training of the neural fingerprinting model did not show an improvement.

This study paves the way for several areas of future research. Investigating alternative methods for aggregating fingerprints that reduce storage while maintaining accuracy is a promising direction. Additionally, evaluating the proposed HRR-based method on larger and more diverse datasets would help assess its scalability and applicability. Furthermore, refining HRR parameters, such as vector dimensionality and the number of position vectors, could enhance the search accuracy.

\bibliographystyle{IEEEtran}
\bibliography{mybib}

\end{document}